# Ramp Reversal Memory and Phase-Boundary Scarring in Transition Metal Oxides

*Naor Vardi, Elihu Anouchi, Tony Yamin, Srimanta Middey, Michael Kareev, Jak Chakhalian, Yonatan Dubi and Amos Sharoni**

N. Vardi, E. Anouchi, T. Yamin, Dr. A. Sharoni
Department of Physics, Bar Ilan University, Ramat-Gan, IL-5290002, Israel
Bar-Ilan Institute of Nanotechnology & Advanced Materials, Ramat-Gan, IL-5290002, Israel
E-mail: amos.sharoni@biu.ac.il

Dr. S. Middey, Dr. M. Kareev, Prof. J. Chakhalian
Department of Physics, University of Arkansas, Fayetteville, Arkansas, 72701, USA

Dr. Y. Dubi
Department of Chemistry, Ben Gurion University, Be'er Sheva, IL-841050, Israel
Ilse-Katz Institute for Nanoscale Science and Technology, Ben-Gurion University of the Negev,Be’er Sheva IL-8410501, Israel

Transition metal oxides (TMOs) are complex electronic systems which exhibit a multitude of collective phenomena. Two archetypal examples are $VO_2$ and $NdNiO_3$, which undergo a metal-insulator phase-transition (MIT), the origin of which is still under debate. Here we report the discovery of a memory effect in both systems, manifest through an increase of resistance at a specific temperature, which is set by reversing the temperature-ramp from heating to cooling during the MIT. The characteristics of this ramp-reversal memory effect do not coincide with any previously reported history or memory effects in manganites, electron-glass or magnetic systems. From a broad range of experimental features, supported by theoretical modelling, we find that the main ingredients for the effect to arise are the spatial phase-separation of metallic and insulating regions during the MIT and the coupling of lattice strain to the local critical temperature of the phase transition. We conclude that the emergent memory effect originates from phase boundaries at the reversal-temperature leaving "scars" in the underlying lattice structure, giving rise to a local increase in the transition temperature. The universality and robustness of the effect shed new light on the MIT in complex oxides.

TMOs are a hallmark example of complex electron systems; the competition between charge, spin, strain, lattice, oxidation and other degrees of freedom, having similar energy scales, give rise to numerous collective phenomena[1-3] including superconductivity,[4] colossal magnetoresistance[5] and metal-insulator transitions (MIT).[6] In many thin film TMOs which exhibit a temperature (T)-driven phase transition, complexity is manifest through the coexistence of multiple phases where a single phase is expected,[7-11] providing the setting required for emergent phenomena to develop.[1]

An intriguing feature found in many of the systems that exhibit the aforementioned phenomena, is the appearance of internal memory, where the system's properties (e.g. resistance or magnetization) depend on the measurement history. Such memory effects appear in various forms and measurement settings, having very different microscopic origins, many of which are still poorly understood. Examples include dynamical memory and slow relaxation in electron-glass systems such as amorphous oxides,[12,13] spatially phase-separated memory in colossal-magnetoresistance manganites,[14] shape-memory in martensitic alloys,[15] and more.[16,17] Here we report the appearance of an unexpected memory effect within the MIT in two prototypical examples of complex TMOs, namely $VO_2$ and $NdNiO_3$ (NNO). The characteristics of the observed effect differ substantially from those of previously reported memory effects, indicating a different microscopic origin.

$VO_2$ and NNO both exhibit an MIT, however its features and microscopic origin are quite different. In $VO_2$ the MIT occurs above room temperature, ~340 K, and is accompanied by a structural transition from low temperature monoclinic to high temperature tetragonal structure. The mechanism is still under debate, showing evidence for electron-electron interaction, i.e. Mott-like transition,[18-20] but also strong coupling to phonon modes and doubling of the unit-cell.[21] In $VO_2$ thin films and nano-beams, the transition develops through a spatially phase separated state[8,10,22] that may affect the dynamics of the transition. Phase separation is complicated by at least one additional insulating phase for which a solid state triple point was identified.[23] Further modification can be achieved by substrate and local strain, which alter the $VO_2$ electronic properties[18] or control the phase separation.[24]

The MIT in NNO displays quite different features. Upon cooling below ~150 K the MIT is accompanied by a simultaneous paramagnetic-to-antiferromagnetic transition, structural symmetry lowering (*Pbnm* to $P2_1/n$) and charge ordering. Phase separation has been recently reported in ultrathin films.[25] Conflicting evidence for the presence of charge ordering[26-29] and the role of electronic vs. magnetic correlations[28,30] point to the intricate nature of this transition (see for example[31] and references therein).

In typical T-driven MIT measurements, the temperature is ramped up from low to high temperatures, driving the system from an insulating to a metallic phase. To observe the memory effect, we perform a series of resistance vs. temperature (R-T) measurements, where we interrupt the usual temperature rise with a reversal of the ramp from heating to cooling at a specific temperature, $T_H$, which lies within the transition region. A regular R-T measurement after the interrupted (ramp-reversed) measurement shows an increase in resistance (compared to the untreated sample), where the resistance increase maximum correlates with $T_H$ (it appears 0.15±0.05 K above $T_H$ in hundreds of measurements with tens of different $T_H$ choices), demonstrating that the sample "remembers" the ramp-reversal temperature. This memory effect is non-volatile (tested after 3 days) and multiple ramp reversals can record multiple temperatures. This "ramp reversal memory" (RRM) appears only for reversing from heating to cooling, and is erased upon increasing the temperature beyond the reversal point.

$VO_2$ thin films were deposited on r-cut sapphire to thickness of 60 nm, with epitaxial growth in the $[2\bar{1}\bar{1}]$ direction. NNO samples were grown epitaxially on $NdGaO_3$ substrate with thickness of 10 nm. All R-T measurements were performed on devices in the 4-probe configuration. **Figure 1** shows the major-loops R-T of $VO_2$ (Figure 1a) and NNO (inset of Figure 1f) as black lines. "Major loops" (ML) refer to R-T measurements having the minimum and maximum temperatures outside of the hysteresis region, i.e. reaching, in principle, a single-phase state of the device. [32] For $VO_2$, this occurs between 310 and 365 K, and for NNO between 50 and 200 K.

We exemplify the ramp reversal sequence in Figure 1b, plotting the temperature vs. time (T-t), rotated to correlate with the T-axis in Figure 1a; the sequence is color coded for clarity. Following one or more ML (black line), the temperature is cycled between a high reversal-temperature ($T_H$) and low reversal-temperature ($T_L$), 10 times in this case, followed by cooling to the low temperature of the ML (purple line). Finally, two ML are performed, green and blue lines in Figure 1b. Figure 1c shows a zoom-in to $T_H$, where we find an unexpected increase in sample resistance (compared to the first ML) at $T_H$ for consecutive cycles and also for the next ML (green), which completely disappears in the following ML (blue). To enable a quantitative analysis, we evaluate the temperature-dependent resistance change between different MLs, marked ΔR in Figure 1d (a further zoom), and plot the normalized resistance change, ΔR/R vs. T. The ΔR/R vs. T plot for the $VO_2$ sequence and for a similar measurement on NNO (see Figure S9 for its T-t sequence) are presented in Figure 1e and 1f, respectively. The green lines show ΔR/R for the first heating ML following the ramp reversal sequence and the blue lines for the consecutive ML. The unambiguous resistance increase, 15% for $VO_2$ and 1.2% for NNO, has a maximum at the reversal temperature (dashed line) in both systems. Also in common is the vanishing of this increase in the next ML.

We note in passing that our measurements allowed us to identify in the $VO_2$ samples glass-related memory effects (see supplementary section S3 for further details). [33] We stress, however, that this effect is not related, but rather coincides, with the RRM effect. To minimize contributions from the glass-dynamics in our measurements we performed fast ramp-reversals cycles (at 5-10 K $min^{-1}$), followed by slower ML (1 K $min^{-1}$) to attain high-resolution and accurate R-T curves.

The central features of the RRM appear in **Figure 2**, where we present a comparison between $\Delta R/R$ from three different ramping sequences (see Figure S9 for their T-t plots): (1) ramping the temperature between $T_L$ and a number of different $T_H$ that decrease with each consecutive cycle ($T_H$ =344, 342 and 340 K for $VO_2$, $T_H$ = 151, 146, 141 and 136 K for NNO), followed by MLs; (2) same as 1 but with increasing $T_H$; and (3) performing identical ramp reversal cycles but with a single reversal temperature. In Figure 2a we show, for a $VO_2$ sample, $\Delta R/R$ following a sequence of decreasing $T_H$ (blue curve) compared with that of increasing $T_H$ (purple curve). Two striking effects can be seen: First, $\Delta R/R$ for decreasing $T_H$ shows three distinguishable peaks, corresponding to the three reversal temperatures. Second, and in contrast, the increasing $T_H$ sequence has only one peak at the final and highest reversal temperature, matching only the high temperature portion of the decreasing sequence. Performing the measurement for NNO results in similar outcomes (Figure 2b): four peaks are found in the $\Delta R/R$ measurement for decreasing $T_H$, but for increasing $T_H$ there is a signature only for the last and highest reversal temperature sequenced. Figure 2c exemplifies, for the $VO_2$ sample, the contribution of cycling to a single reversal temperature relative to the three-$T_H$ sequence. Each individual cycle has a resolvable resistance peak corresponding to its own $T_H$. Moreover, a simple weighted sum of the individual curves reconstructs the three-$T_H$ result (see legend for details). All features were very well reproduced, see Figure S12 for examples. These results, put together, establish that (1) the ramp reversal sequence “encrypts” in the TMOs a “memory” of the reversal temperature, manifested as a resistance increase at $T_H$; thus we term the effect “ramp reversal memory” – RRM, (2) It is possible to write and resolve a number of memory events, but (3) increasing the temperature further erases all memories set at lower temperatures, and (4) the same phenomena appears in two very different TMOs, with different mechanisms for the T-driven MITs; notably, NNO has an antiferromagnetic component to the transition while $VO_2$ has no magnetic element.

To further characterize the RRM effect and resolve the microscopic mechanism, we performed additional measurements on the $VO_2$ samples (T-t procedures are presented in Figure S10 and S11): (I) **RRM in cooling MLs**: **Figure 3**a shows the resistance difference ΔR/R along a cooling ML, performed after the sample was heated to the metallic state following a ramp reversal protocol. The small value of ΔR/R (within the measurement noise level) demonstrates that there is no RRM for the cooling sequence, i.e. the high temperature metallic phase erases the memory. (II) **Durability of RRM at low T**: We tested whether the RRM is erased by cooling to low temperatures. Following a reversal cycle at $T_H$ = 344 K, the sample was cooled to 100 K, deep in the insulating regime. A subsequent measurement of ΔR/R along the ML (Figure 3b) clearly reveals a peak in ΔR/R at $T_H$, indicating that, in contrast to heating, cooling does not erase the memory. (III) **Stability of the RRM**: the time dependence of the memory effect was tested by holding the sample at 310 K for 16 hours following a ramp reversal sequence. Comparing the resistance change to a sequence without the ramp reversal (but similar dwell time), we find that the memory effect remains intact after 16 hours (Figure 3c). We note that the dwell time is longer than the relaxation time of the sample in the glassy state (inferred from Figure S5). (IV) **Magnitude of the RRM effect**: The effect of different number of reversal cycles is shown in Figure 3d. The maximum in ΔR/R ($\Delta R/R_{max}$) is plotted vs. number of cycles ($N_{cycles}$), showing that the maximum amplitude increases, and saturates exponentially with the number of cycles, the purple line is a fit to an exponential decay (full ΔR/R vs. T are provided in Figure S13). We further find that the predominant factor determining the magnitude of the RRM effect is the sample's temperature coefficient of resistance (TCR), $\frac{dR/dT}{R}$. In Figure 3e we plot the maximal resistance change after a single ramp reversal at $T_H$, as a function of $T_H$ (black circles, the line is a guide to the eye). On the same figure we plot the TCR as a function of temperature (purple line). Both the maximal $\Delta R/R$ and the TCR exhibit a similar line-shape, with a maximum at $T_C$~341 K. Finally, in Figure 3f we plot how $\Delta R/R_{max}$ behaves upon changing $T_L$ (for $T_H$ = 347 K). While $T_L$ < 335 K, $\Delta R/R_{max}$ is nearly constant, after which it decreases linearly to zero. Using the effective medium approximation[34] we find that the slope-change point at 335 K roughly coincides with the percolation threshold of the phase separated system.[10]

We stress there are fundamental differences between the RRM properties and other memory effects in TMOs. First, it is not of magnetic origin, as opposed to the memory reported in manganite systems or other magnetic-particle systems, as is clear from the fact that $VO_2$ is non-magnetic. Second, its non-volatile nature is in contrast to the time-dependence of the dynamical glass memory. Third, in contrast to, e.g., shape-memory alloys, here it is possible to encode in the memory a specific or even several specific reversal temperatures. Finally, we have not performed any drastic manipulation at the reversal temperature, like a high electric field or light pulse, that can alter the state of the system, e.g. through oxygen motion or heating.[17,35]

The features displayed in Figure 2 and 3 provide hints onto the origin of the RRM. We propose a microscopic mechanism that qualitatively captures all of the features. We recall that the temperature driven MIT occurs through phase coexistence,[8,10,22,32,36] i.e. during heating metallic domains nucleate and then grow, and vice-versa during cooling. We hypothesize that reversing of the temperature ramp results in a modification of the local properties of the sample – and specifically the local transition temperature – at the phase-boundaries between insulating and metallic domains. Put simply, we posit that when the temperature ramp is reversed, the direction change of the phase-boundary motion creates a “scar”- a static one dimensional deformation of the lattice along the phase boundary, which has an elevated local critical temperature.

To see how this leads to RRM, in **Figure 4**a we provide a cartoon that schematically portrays its development. When heating from low temperatures, metallic domains appear and grow (a1, phase boundary is marked). Upon ramp reversal at $T_H$, the metallic domains start to shrink, and a scar develops at the phase boundary with an increased local $T_C$ (a2). Upon reheating (a3), the metallic domain cannot easily cross the scarred boundaries. Additional metallic domains may nucleate and grow, having their own phase boundaries. Upon the next reversal at $T_H$ the sample acquires excess scarring (a4). If we heat to a high enough temperature $T'_H$ (a5), the metallic domains will eventually continue to grow, cross the scarred boundaries and "heal" the scars, by that erasing the memory. Finally, cooling the system from this $T'_H$ results in new scars that imprint the new reversal temperature. If we start the process with a metallic film, and insulating phases appear, upon re-heating, the metallic domains grow and no scars are formed at the phase boundaries.

Figure 4b - 4d show the results of numerical calculations performed based on our model for $VO_2$ (see methods and section S2 in supplementary information for details). Only two assumptions enter the calculation. First, we introduce a spatially correlated distribution of the local MIT transition temperature, realizing experimental reports of the spatially dependent phase transition[8] and the appearance of nucleation centers of the metallic phase. Figure 4b presents an example for such a realization, where the local $T_C$ (color coded) is randomly distributed in the lattice (size 40x40 for this example), the distribution centered on the transition temperature and the width of the distribution defining the width of the transition.[32] The second assumption is our postulate, i.e. upon reversing T-ramp from heating to cooling, $T_C$ increases along the spatially separated phase boundaries. Figure 4c depicts local changes in $T_C$ (the "scars") following three cycles with decreasing $T_H$ (values for $T_H$ same as in Figure 2a). In Figure 4d we plot $\Delta R/R$, numerically calculated for the same protocol as Figure 2c, i.e. one sequence with three different and decreasing values of $T_H$ (blue curve), compared to three cycles with a single $T_H$ (red curves). A weighted sum of the three cycles is plotted as a dashed red line. The numerical results show remarkable resemblance to the memory effect as presented in Figure 2 (compare to Figure 2c).

This model qualitatively captures the RRM properties presented in Figure 1 and 3, and specifically the different factors determining the magnitude of the RRM. First, the difference in ΔR/R magnitude between $VO_2$ and NNO is related to the TCR of $VO_2$ being much larger than NNO. Since resistance change come from an increase in local $T_C$ of boundaries, our model implies that ΔR/R should be larger in $VO_2$, as evidenced in Figure 1e and 1f. The ratio of TCRs for $VO_2$ and NNO is ~10 (-0.68 for $VO_2$ and -0.054 for NNO at the measured temperatures), similar to the RRM ratio. In a similar manner, changing $T_H$ should follow the change in TCR, as evidenced in Figure 3e. Furthermore, performing more reversal cycles will introduce more nucleation opportunities and more scarring, up to the stage where each cycle adds the same number of scars as it heals, corresponding with the observed features of Figure 3d. The results of Figure 3f can be understood by noting that cycling to higher $T_L$ reduces the probability of additional metallic nucleating cites and domains (see Figure 4a), leading to less phase boundaries and a smaller ΔR/R. Interestingly, the signal starts decreasing only for $T_L$ above the percolation threshold, since above the percolation temperature metallic domains remain connected, thus less scarring can occur.

A plausible origin for the formation of scars involves a strain-stress field distribution that develops at the boundaries between the spatially separated metallic and insulating phases due to the ramp reversal, stabilized by the underlying substrate interface.[37] This implies that the RRM is limited to thin films, and will be absent in bulk samples. The effect of strain fields can result in an increased $T_C$ of the boundaries.[24,38,39] The width of the scarred phase-boundary must be smaller than the size of the phase-separated regions, which may be a number of nanometers in our TMOs.[8,11] In principle the scarred phase boundaries may be only a few unit-cells wide. This can explain the lack of interference we find in the RRM when several $T_H$ are set simultaneously. In general, there is nothing inherent dictating that the memory should be erased by the metallic high-temperature phase, but in such a case the strain fields have to result in a $T_C$ decrease for one to observe the memory effect. The remarkable correlation between $T_H$ and the RRM maximum (see Figures. S12 and S13) indicates that the strain fields' energy are strongly correlated with the reversal temperature.

The appearance of the RRM effect in two systems with very different MIT mechanisms and the ability of a minimal model to capture the essential features of the effect, allow us to predict that the RRM is universal, and should appear in other thin-film systems which exhibit first order phase transitions and have several meta-stable states of the same phase. Moreover, the RRM should not be limited to a temperature ramp, and may appear with any transition-driving external force. This opens a new route to study emergent phenomena in metal oxides specifically and in complex systems in general, and may lead to future memory-based applications utilizing this multi-state non-volatile effect.

**Experimental Section**

*Film Deposition*: $VO_2$ films, 65nm thick, were deposited on R-cut $(1\bar{1}02)$ sapphire substrates by reactive RF magnetron sputtering in an $Ar/O_2$ mixture from a nominal $V_2O_3$ target. Optimal growth occurred at a substrate temperature of 600 °C, total pressure of 3 mtorr, oxygen partial pressure of 2% and a growth rate of 0.17 Å $s^{-1}$. Under these conditions the films grew smoothly, epitaxial and single-phased, as confirmed by AFM, X-ray diffraction and HR-SEM (see also Figure S1). More details and characteristics can be found in the supplementary material and previous work.[40] We present measurements of two configuration of $VO_2$ samples. In one sample wire bonding electrodes was connected directly to the film. The other was patterned to a 10 μm wide and 100 μm long channel with vanadium electrodes. Both showed identical RRM behavior.

Epitaxial thin film (thickness ~10 nm) of $NdNiO_3$ was stabilized on a single crystalline $NdGaO_3$ $(001)_{pc}$ substrate (*pc= pseudocubic*) at 680C under dynamic oxygen pressure of 150

mTorr using the pulsed laser interval deposition method.[41,42] The layer-by-layer growth (Figure S2) was monitored by *in-situ* high pressure RHEED (reflection high energy electron diffraction). The film was post-annealed *in-situ* in one atmosphere of $O_2$ for 30 minutes at the growth temperature. X-ray diffraction measurement confirmed appropriate film-substrate epitaxy and the desired +3 oxidation state of Ni was verified by synchrotron based X-ray absorption spectroscopy.[41,42]

*Transport Measurements*: R-T measurements were acquired in three different cryostats (commercial QD-PPMS, commercial Janis closed cycle refrigerator and home-made insert), all showing similar results, via a four-probe geometry. We used a constant current source to avoid excess Joule heating when the transition to the metallic phase begins. The R-T measurements were performed while continuously sweeping the temperature at controlled rates. We tested a wide range of ramp rates to confirm reliability of temperature sensor and sample temperature, and to ensure reproducibility.

*Numerical Modeling:* The resistance is calculated from a resistor network (RN) model (further details, including all numerical parameters, can be found in section S2 of supplementary). The RN is constructed as follows. First, we construct square lattice of sites which are defined as “insulating” (I) or “metallic” (M) depending on whether the local critical temperature $T_C(r)$ is higher or lower than the temperature T. $T_C(r)$ is extracted from a normal distribution with local correlations, to mimic the appearance of nucleation center in the metallic phase upon crossing the transition temperature (an example of a 40x40 lattice is depicted in Figure 3c). Next, the conductance of each bond of the square lattice are determined, depending on whether it connects two I-sites, two M-sites or one I and one M-sites,

$$G_{I-I} = G_R \exp\left(-\frac{T_0}{T}\right), G_{I-M} = G_{M-M} = R_M^{-1} \quad (1)$$

, where $R_M$ is the resistance (per square) of the sample in the metallic phase, $G_R$ is a constant that characterizes the resistance of the insulating phase, and $T_0$ is the activation gap of the insulating phase. The resistance of the entire RN is then evaluated using Kirchhoff's laws (SI.6).

To account for the "scarring" at temperature ramp reversal, the simulation follows the temperature protocol, and at the ramp reversal $T_C(r)$ is modified such that if an M-site is neighboring an I-site, the local critical resistance of the I-site is increased by $\delta T_C$. Note that all other parameters (i.e. the resistances at the I and M phases and the parameters related to the distribution of $T_C(r)$) can essentially be determined from the major loop measurement,[32] leaving $\delta T_C$ as the only free parameter used to obtain Figure 3c.

**Supporting Information**
Supporting Information is available online from the Wiley Online Library or from the author.

**Acknowledgements**
A.S was supported by the ISRAEL SCIENCE FOUNDATION (grant No. 569/16), S.M. and M.K. were supported by the DOD-ARO under Grant No. 0402-17291. J. C. was supported by the Gordon and Betty Moore Foundations EPiQS Initiative through Grant No. GBMF4534. We thank A. Berger for discussions at the early stage of this research, Y. Yeshurun and D. Ben-Basat for discussions and comments.

Received: ((will be filled in by the editorial staff))
Revised: ((will be filled in by the editorial staff))
Published online: ((will be filled in by the editorial staff))

[1] E. Dagotto, *Science* **2005**, *309*, 257.

[2] J. Ngai, F. Walker, C. Ahn, *Annu. Rev. Mater. Res.* **2014**, *44*, 1.

[3] S. Dorogovtsev, A. Goltsev, J. Mendes, *Rev. Mod. Phys.* **2008**, *80*, 1275.

[4] J. Bednorz, K. Müller, *Z. Phys. B Con. Mat.* **1986**, *64*. 189. doi:10.1007/BF01303701

[5] H. Kuwahara, Y. Tomioka, Y. Moritomo, A. Asamitsu, M. Kasai, R. Kumai, Y. Tokura, *Science* **1996**, *272*, 80.

[6] F. Morin, *Phys. Rev. Lett.* **1959**, *3*, 34.

[7] M. Fäth, S. Freisem, A. Menovsky, Y. Tomioka, J. Aarts, J. Mydosh, *Science* **1999**, *285*, 1540.

[8] M. Qazilbash, M. Brehm, B. Chae, P. Ho, G. Andreev, B. Kim, S. Yun, A. Balatsky, M. Maple, F. Keilmann, H. Kim, D. Basov, *Science* **2007**, *318*, 1750.

[9] E. Dagotto, T. Hotta, A. Moreo, *Physics Reports* **2001**, *344*, 1.

[10] A. Sharoni, J. Ramírez, I. Schuller, *Phys. Rev. Lett.* **2008**, *101*, 26404. DOI 10.1103/physrevlett.101.026404.

[11] G. Mattoni, P. Zubko, F. Maccherozzi, A. van der Torren, D. Boltje, M. Hadjimichael, J. Aarts, *arXiv preprint arXiv:1602.04445, **2016***.

[12] A. Vaknin, Z. Ovadyahu, M. Pollak, *Phys. Rev. B* **2002**, *65*, 134208. DOI 10.1103/physrevb.65.134208.

[13] A. Amir, Y. Oreg, Y. Imry, *Annu. Rev. Condens. Matter Phys.* **2011**, *2*, 235.

[14] P. Levy, F. Parisi, L. Granja, E. Indelicato, G. Polla, *Phys. Rev. Lett.* **2002**, *89*, 137001. DOI 10.1103/physrevlett.89.137001.

[15] J. Zhang, X. Ke, G. Gou, J. Seidel, B. Xiang, P. Yu, W. Liang, A. Minor, Y. Chu, G. Van Tendeloo, X. Ren, R. Ramesh, *Nat. Commun.* **2013**, *4*, 3768. DOI 10.1038/ncomms3768.

[16] A. Sawa, *Materials Today* **2008**, *11*, 28.

[17] J. Jeong, N. Aetukuri, T. Graf, T. Schladt, M. Samant, S. Parkin, *Science* **2013**, *339*, 1402.

[18] N. Aetukuri, A. Gray, M. Drouard, M. Cossale, L. Gao, A. Reid, R. Kukreja, H. Ohldag, C. Jenkins, E. Arenholz, K. Roche, H. Dürr, M. Samant, S. Parkin, *Nat Phys* **2013**, *9*, 661.

[19] A. Gray, J. Jeong, N. Aetukuri, P. Granitzka, Z. Chen, R. Kukreja, D. Higley, T. Chase, A. Reid, H. Ohldag, M. Marcus, A. Scholl, A. Young, A. Doran, C. Jenkins, P. Shafer, E. Arenholz, M. Samant, S. Parkin, H. Dürr, *Phys. Rev. Lett.* **2016**, *116*, 116403. DOI 10.1103/physrevlett.116.116403.

[20] V. Morrison, R. Chatelain, K. Tiwari, A. Hendaoui, A. Bruhacs, M. Chaker, B. Siwick, *Science* **2014**, *346*, 445.

[21] J. Budai, J. Hong, M. Manley, E. Specht, C. Li, J. Tischler, D. Abernathy, A. Said, B. Leu, L. Boatner, R. McQueeney, O. Delaire, *Nature* **2014**, *515*, 535.

[22] C. Yee, L. Balents, *Phys. Rev. X* **2015**, *5*, 021007. DOI 10.1103/physrevx.5.021007.

[23] J. Park, J. Coy, T. Kasirga, C. Huang, Z. Fei, S. Hunter, D. Cobden, *Nature* **2013**, *500*, 431.

[24] J. Cao, E. Ertekin, V. Srinivasan, W. Fan, S. Huang, H. Zheng, J. Yim, D. Khanal, D. Ogletree, J. Grossman, J. Wu, *Nat. Nanotech* **2009**, *4*, 732.

[25] A. Hauser, E. Mikheev, N. Moreno, T. Cain, J. Hwang, J. Zhang, S. Stemmer, *Appl. Phys. Lett.* **2013**, *103*, 182105.

[26] D. Meyers, S. Middey, M. Kareev, J. Liu, J. Kim, P. Shafer, P. Ryan, J. Chakhalian, *Phys. Rev. B* **2015**, *92*, 235126. DOI 10.1103/physrevb.92.235126.

[27] M. Hepting, M. Minola, A. Frano, G. Cristiani, G. Logvenov, E. Schierle, M. Wu, M. Bluschke, E. Weschke, H. Habermeier, E. Benckiser, M. Le Tacon, B. Keimer, *Phys. Rev. Lett.* **2014**, *113*, 227206. DOI 10.1103/physrevlett.113.227206.

[28] M. Upton, Y. Choi, H. Park, J. Liu, D. Meyers, J. Chakhalian, S. Middey, J. Kim, P. Ryan, *Phys. Rev. Lett.* **2015**, *115*, 036401. DOI 10.1103/physrevlett.115.036401.

[29] D. Meyers, J. Liu, J. Freeland, S. Middey, M. Kareev, J. Kwon, J. Zuo, Y. Chuang, J. Kim, P. Ryan, J. Chakhalian, *Sci. Rep.* **2016**, *6*.

[30] M. Stewart, J. Liu, M. Kareev, J. Chakhalian, D. Basov, *Phys. Rev. Lett.* **2011**, *107*, 176401. DOI 10.1103/physrevlett.107.176401.

[31] S. Middey, J. Chakhalian, P. Mahadevan, J. Freeland, A. Millis, D. Sarma, *Annu. Rev. Mater. Res.* **2016**, *46*, 305. DOI: 10.1146/annurev-matsci-070115-032057.

[32] J. Ramírez, A. Sharoni, Y. Dubi, M. Gómez, I. Schuller, *Phys. Rev. B* **2009**, *79*, 235110. DOI 10.1103/physrevb.79.235110.

[33] V. Dobrosavljević, D. Tanasković, A. Pastor, *Phys. Rev. Lett.* **2003**, *90*, 016402. DOI 10.1103/physrevlett.90.016402.

[34] T. Yamin, Y. Strelniker, A. Sharoni, *Sci. Rep.* **2016**, *6*, 19496.

[35] S. Bae, S. Lee, H. Koo, L. Lin, B. Jo, C. Park, Z. Wang, *Adv. Mater.* **2013**, *25*, 5098.

[36] G. Catalan, *Phase Transitions* **2008**, *81*, 729.

[37] H. Hwang, Y. Iwasa, M. Kawasaki, B. Keimer, N. Nagaosa, Y. Tokura, *Nat. Mat.* **2012**, *11*, 103.

[38] J. Sohn, H. Joo, K. Kim, H. Yang, A. Jang, D. Ahn, H. Lee, S. Cha, D. Kang, J. Kim, M. Welland, *Nanotechnology* **2012**, *23*, 205707.

[39] L. Pellegrino, M. Biasotti, E. Bellingeri, C. Bernini, A. Siri, D. Marré, *Adv. Mater.* **2009**, *21*, 2377.

[40] T. Yamin, T. Havdala, A. Sharoni, *Mater. Res. Expr.* **2014**, *4*, 046302.

[41] J. Liu, M. Kareev, B. Gray, J. Kim, P. Ryan, B. Dabrowski, et al. *Appl. Phys. Lett.* **2010**, *96*, 233110.

[42] J. Liu, M. Kargarian, M. Kareev, B. Gray, P. Ryan, A. Cruz, et al. *Nat. Commun.* **2013**, *4*, 2714.

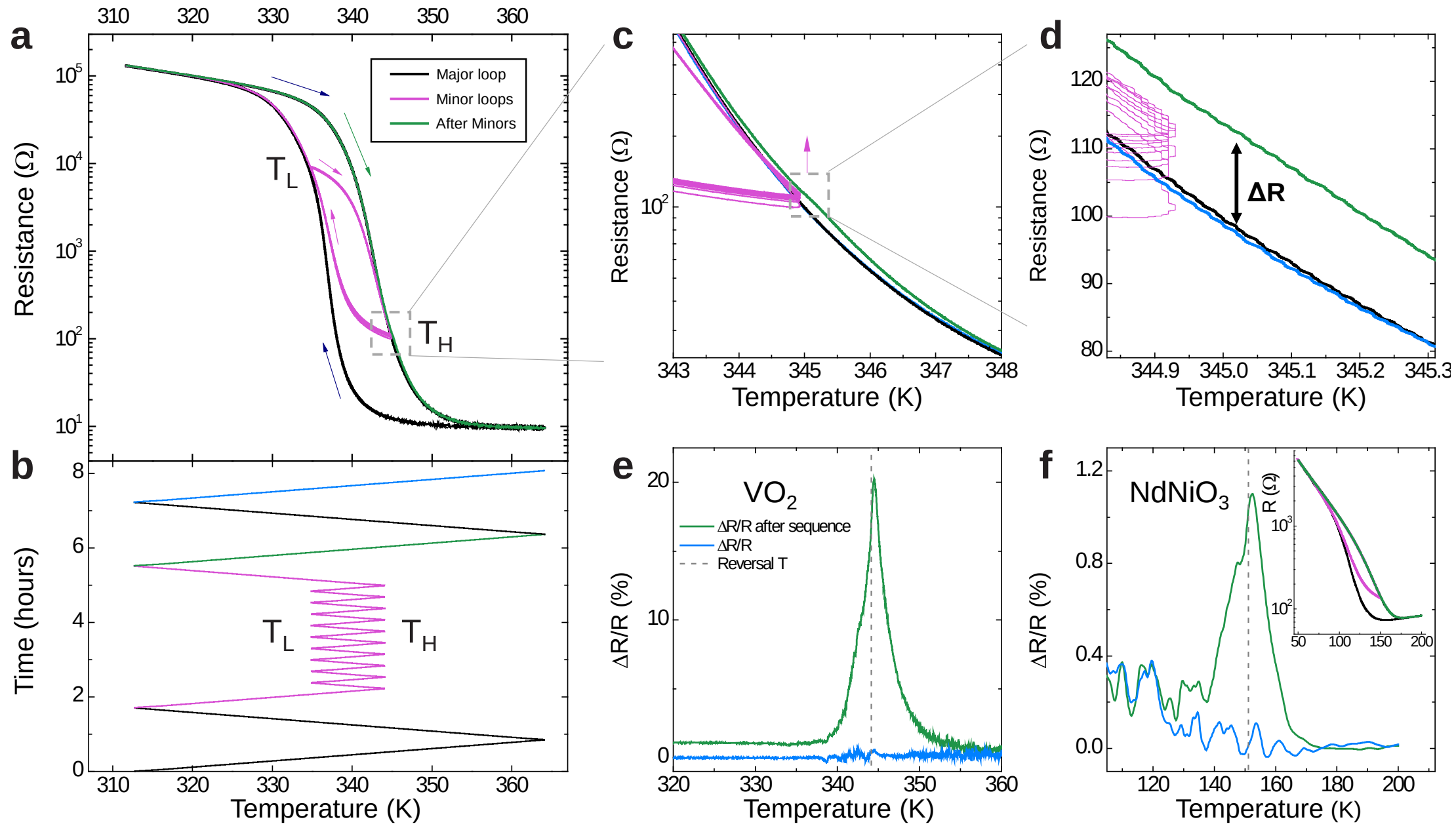

**Figure 1.** Ramp reversal sequence. a) R-T curve of full ramp reversal sequence and (b), corresponding temperature ramps, $T_H$ and $T_L$ are marked in both. The different parts of the sequence are color coded; black- major loops, purple- reversal loops, green and blue- 1st and 2nd heating curves following reversal loops. c,d) Zoom-ins of the $T_H$ region, arrow marks sequential reversal loops, ΔR is marked in (d). e,f) plots of ΔR/R for $VO_2$ (e) and NNO (f), for the 1st (green) and 2nd (blue) heating curves. Inset f) R-T of NNO, using the same color coding as in (a).

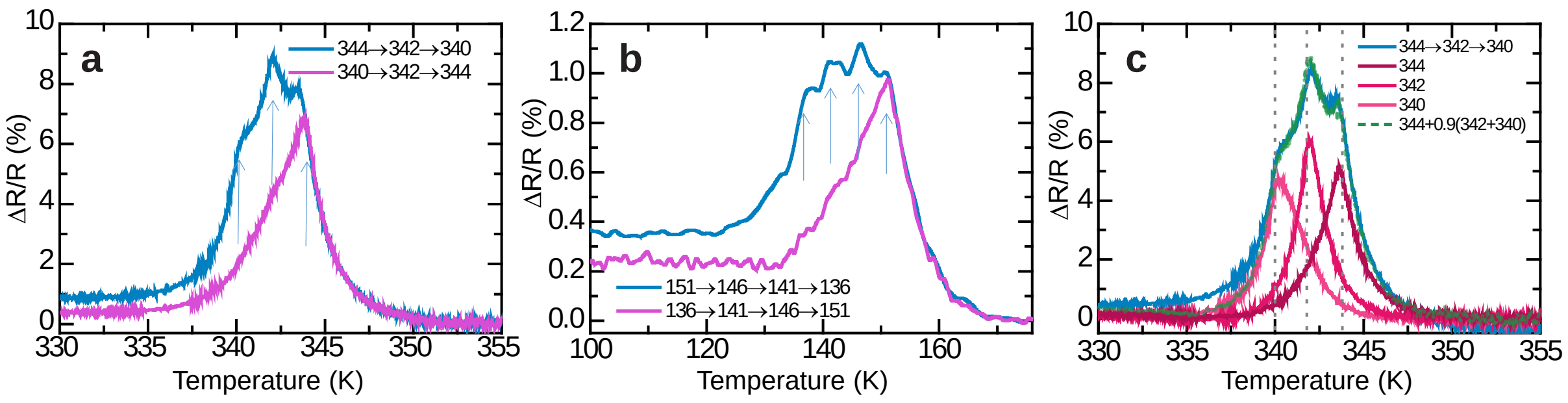

**Figure 2.** Ramp reversal memory main properties. a) ΔR/R in $VO_2$ from sequence with three different reversal temperatures, $T_H$=340,342 and 344 K. Green- decreasing $T_H$ between cycles, blue increasing $T_H$. b) Same as (a), but for NNO and with four $T_H$, $T_H$=136, 141, 146 and 151 K. c) Single reversal loops for same temperatures and their weighted sum compared to the three $T_H$ sequence, see legend.

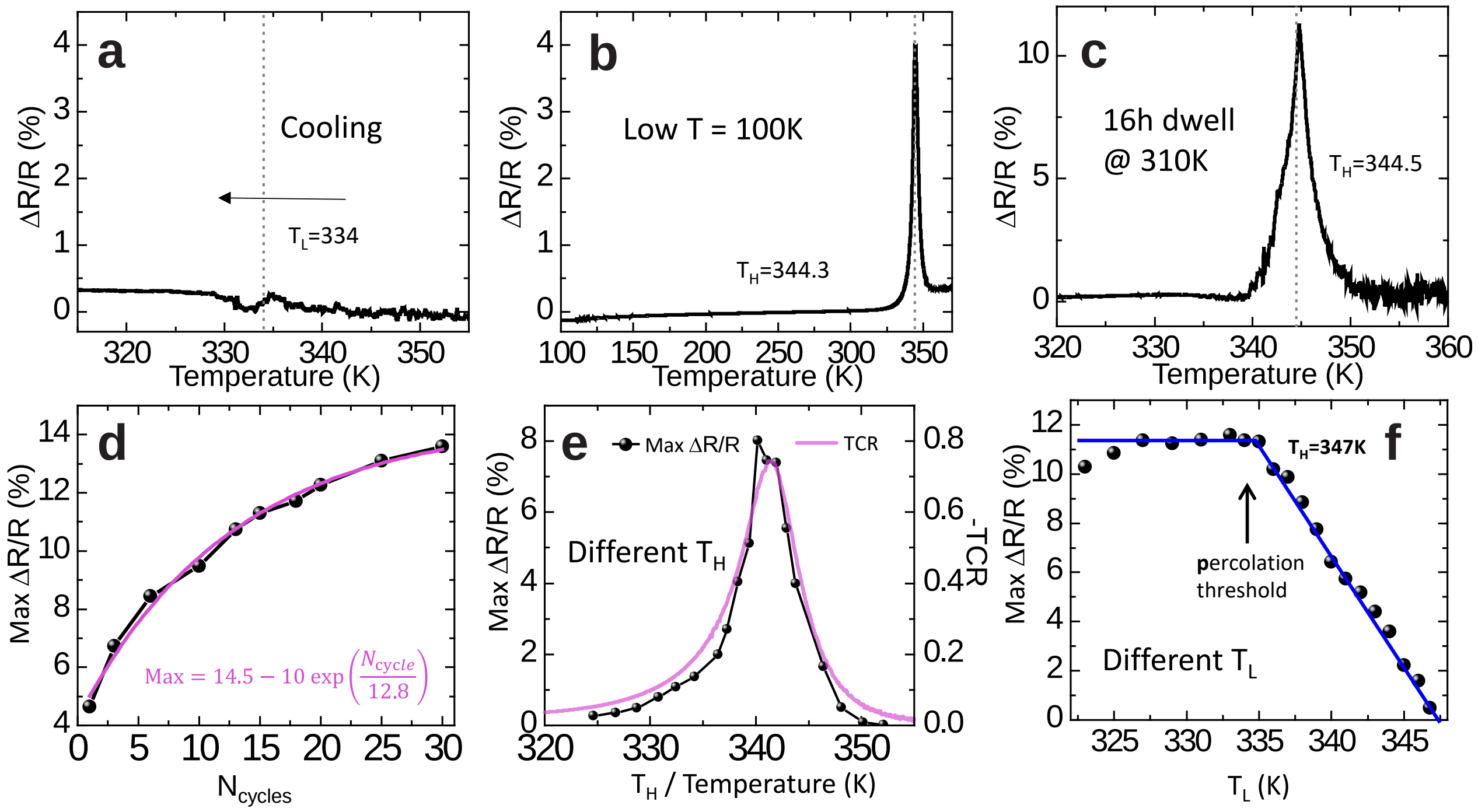

**Figure 3.** Ramp reversal memory characteristics demonstrated for $VO_2$. a) ΔR/R for cooling ML. b) ΔR/R after cooling to 100K. c) ΔR/R following a 16 hour dwell at 310K. d) $\Delta R/R_{Max}$ vs. number of reversal cycles, black circles. Purple line is a fit to exponential decay. e) Peak of ΔR/R as a function of $T_H$ - black circles and left axis, and TCR of $VO_2$ sample for same heating curve - purple line and right axis. f) Maximum of ΔR/R vs. $T_L$ for $T_H$=347 K. Blue solid line is a guide to the eye.

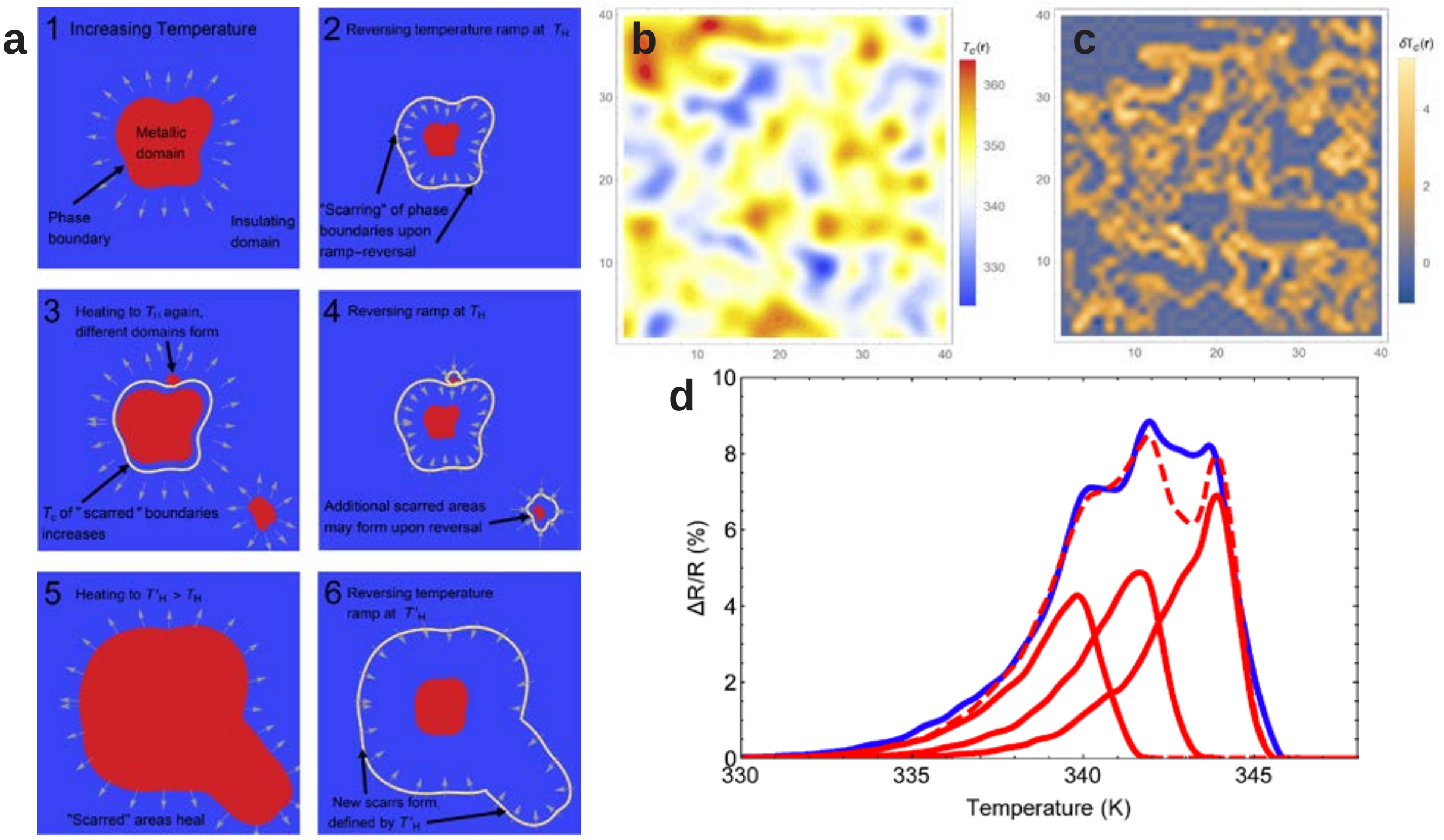

**Figure 4.** RRM model. a) Cartoon of ramp reversal memory showing scarring during heating and cooling cycles, explanation appears in text. Blue- insulating, red- metallic, white- scarred phase boundaries. b) Numerical realization of spatial fluctuation in $T_C$. c) Spatial mapping of local $T_C$ change due to numerical realization of three reversal temperatures 344, 342 and 340 K. d) Simulation of ΔR/R for decreasing $T_H$ between each cycle (blue) and for three single reversal loops (red). Dashed red line is the sum of three single reversal loops.